\documentclass[aps,prl,twocolumn,superscriptaddress,showpacs]{revtex4-2}
\usepackage{amsmath, amssymb,wasysym,graphicx}
\usepackage{appendix}
\usepackage[ruled]{algorithm2e}
\usepackage{lineno}

\usepackage[utf8]{inputenc}
\usepackage[T1]{fontenc}
\usepackage{xcolor}

\usepackage{diagbox}
\usepackage{tabularx}

\IfFileExists{newtxtext.sty}
{\usepackage{newtxtext,newtxmath}}
{\IfFileExists{stix.sty}
{\usepackage{stix}}
{\IfFileExists{mathptmx.sty}
{\usepackage{mathptmx}}{} } }

\usepackage{textcomp}

\usepackage{bm}

\usepackage{booktabs,siunitx}

\pdfoutput=1
\usepackage{color}
\definecolor{LinkColor}{rgb}{0.256,0.439,0.588}
\usepackage{hyperref}
\hypersetup{
colorlinks=true,
citecolor=LinkColor,
linkcolor=LinkColor,
urlcolor=LinkColor
}

\usepackage{multirow}

\begin{document}
%\linenumbers 
\title{The teaching from entanglement: 2D SU(2) antiferromagnet to valence bond solid deconfined quantum critical points are not conformal}

\author{Yuan Da Liao}
\email{ydliao@fudan.edu.cn}
\affiliation{State Key Laboratory of Surface Physics, Fudan University, Shanghai 200438, China}
\affiliation{Center for Field Theory and Particle Physics, Department of Physics, Fudan University, Shanghai 200433, China}

\author{Gaopei Pan}
\email{gppan@iphy.ac.cn}
\affiliation{Beijing National Laboratory for Condensed Matter Physics and Institute of Physics, Chinese Academy of Sciences, Beijing 100190, China}
\affiliation{School of Physical Sciences, University of Chinese Academy of Sciences, Beijing 100049, China}

\author{Weilun Jiang}
\affiliation{Beijing National Laboratory for Condensed Matter Physics and Institute of Physics, Chinese Academy of Sciences, Beijing 100190, China}
\affiliation{School of Physical Sciences, University of Chinese Academy of Sciences, Beijing 100049, China}

\author{Yang Qi}
\email{qiyang@fudan.edu.cn}
\affiliation{State Key Laboratory of Surface Physics, Fudan University, Shanghai 200438, China}
\affiliation{Center for Field Theory and Particle Physics, Department of Physics, Fudan University, Shanghai 200433, China}
\affiliation{Collaborative Innovation Center of Advanced Microstructures, Nanjing 210093, China}

\author{Zi Yang Meng}
\email{zymeng@hku.hk}
\affiliation{Department of Physics and HKU-UCAS Joint Institute
of Theoretical and Computational Physics, The University of Hong Kong,
Pokfulam Road, Hong Kong SAR, China}

\begin{abstract}
The deconfined quantum critical point (DQCP) -- the enigmatic incarnation of the quantum phase transition beyond the Landau-Ginzburg-Wilson paradigm of symmetries and their spontaneous breaking -- has been proposed and actively pursued for more than two decades~\cite{senthilDeconfined2004,senthilQuantum2004,qinDuality2017,wangDeconfined2017}. Various 2D quantum many-body lattice models, both in spin/boson~\cite{sandvikEvidence2007,louAntiferromagnetic2009,sandvikContinuous2010} and fermion~\cite{liuSuperconductivity2019,liaoDiracI2022} representations have been tested with the state-of-the-art numerical techniques and field-theoretical analyses, and yet, the conclusion is still controversial. Experimental realizations of DQCP in the quantum magnet SrCu$_2$(BO$_3$)$_2$~\cite{zayed4spin2017,guoQuantum2020, jimenezquantum2021} and superconducting quantum criticality in 2D material have either shown first order transition or intermediate phase. The tension between the lattice scale details and the requirement from continuum limit, manifested in the form of the inconsistent critical scaling behavior and violations of generic conformal bootstrap bound, has not been resolved. 
Here we solve these decades-long controversies from the new and fundamental perspective of the quantum entanglement. We develop the incremental algorithm to compute the entanglement entropy at a fermionic DQCP with unprecedentedly accurate data and reveal the universal coefficient of the logarithmic correction therein is negative and at odds with positivity requirement of the conformal field theory. Together with results in other 2D DQCP lattice models (both in fermion and spin systems)~\cite{zhaoScaling2022,wangScaling2021,liuFermion2022}, our discoveries clearly demonstrate the 2D SU(2) antiferromagnet to valence bond solid DQCPs are not conformal fixed point and naturally explain the experimental difficulties in finding them. This marks the end of the beginning of unambiguous finding of the quantum phase transitions truely beyond the Landau-Ginzburg-Wilson paradigm, since its suggestion two decades ago.
\end{abstract}
\date{\today}
\maketitle

\section{Introduction} 
The development of the deconfined quantum critical point (DQCP) has been enigmatic. In the beginning, the proposal of DQCP is meant to offer an alternative of quantum critical point beyond the symmetry breaking paradigm of Landau-Ginzburg-Wilson~\cite{senthilQuantum2004,senthilDeconfined2004}. However, since the discovery of the first concrete 2D lattice model -- the spin-1/2 SU(2) JQ model~\cite{sandvikEvidence2007,louAntiferromagnetic2009,sandvikContinuous2010} --  which offered evidence of hosting a DQCP between antiferromagnetic (AFM) N\'eel state and valence bond solid (VBS) state from quantum Monte Carlo (QMC) simulations, the discussion and the focus of the field gradually shift from verifying %with the most advanced numerical and analytical approaches 
whether the proposed lattice models realized the DQCP with expected properties such as the emergent continuous symmetry~\cite{nahumEmergent2015,nahumDeconfined2015,zhangContinuous2018} and $\mathbb{C} \mathbb{P}^{N-1}$ fixed points~\cite{blockFate2013,dEmidioNew2017,xuMonte2019}, to explore the new phenomena beyond the original scope such as the two length scales~\cite{shaoQuantum2016}, the duality between DQCP and bosonic topological transition~\cite{qinDuality2017,wangDeconfined2017,shuNonequilibrium2022,shuDual2022}, the emergent Dirac spinon coupled with dynamic gauge fields~\cite{maDynamical2018,maRole2019,wangDynamics2019,janssenConfinement2020} and the weakly first-order, multiplicity and complex fixed point of the transition~\cite{nahumNote2020,maTheory2020,zhaoMulticritical2020,kuklovDeconfined2008,jiangFrom2008,chenDeconfined2013,dEmidioDiagnosing2021}, to name a few. It seems that the exploration of the properties of lattice models, even if they turn out to defy the original proposal of DQCP, have become a lively research direction on its own right, and such activities attract broad interests across computational, high-energy and field-theoretical~\cite{nakayamNecessary2016,liBootstrapping2022} to experimental~\cite{zayed4spin2017,guoQuantum2020,sunEmergent2021, jimenezquantum2021,cuiProximate2022} physics communities and the collective efforts have formed into a very coherent stream.

Within such collective activities, there emerges different classes of lattice models besides the JQ type which might host a DQCP. These models include 1D spin systems~\cite{robertsDeconfined2019,huangEmergent2019}, and more importantly, 2D interacting Dirac fermion systems with VBS-AFM transition~\cite{liaoDiracI2022,liaoDiracIII2022,satoDirac2017,liFermion2017,liDeconfined2019} or quantu spin Hall (QSH) to superconductivity (SC) transition~\cite{liuMetallic2022,liuSuperconductivity2019,wangDoping2021,wangPhases2021}. More recently, there is a new proposal of a conformal DQCP with SO(5)$_f \times$ SO(5)$_b$ global symmetry that describes the cuprate phase diagram with pseudogap metal, d-wave superconductivity and charge order as various symmetry breaking phases~\cite{christosModel2023}. Different from the spin models such as the JQ, these fermionic lattice models enjoy their own advantages such as the absence of symmetry allowed quadruple monopoles and the associated second length scale which breaks the assumed U(1) symmetry down to  $\mathbb{Z}_4$~\cite{liuMetallic2022,liuSuperconductivity2019,wangDoping2021,wangPhases2021,liaoDiracI2022,shaoQuantum2016}, and the satisfaction of the conformal Bootstrap bound imposed on critical exponents for emergent continuous symmetry~\cite{liaoDiracI2022,nakayamNecessary2016}. Therefore, even if the determinant QMC (DQMC) simulations for fermion models are much more expensive compared with that of the stochastic series expansion QMC for spin models (computational complexities of the former $\sim O(\beta N^3)$ where the inverse temperature $\beta=\frac{1}{T}$ and system size $N=L^d$ with $d$ the spatial dimension, compared with $\sim O(\beta N)$ for the latter), the computation and the associated theoretical exploration of the 2D fermionic DQCP models flourish~\cite{liuFermion2022,satoSimulation2022,liaoDiracI2022} nevertheless.

Experimentally, both the VBS-AFM and the QSH-SC types of DQCP transitions have been actively pursued in quantum magnets SrCu$_2$(BO$_3$)$_2$~\cite{zayed4spin2017,guoQuantum2020, jimenezquantum2021,cuiProximate2022} and alike and in the superconducting quantum criticality in various 2D materials, but the results till now show that at the suggested DQCP, the systems either exhibit first order transition or an intermediate phase, and thence defied the original proposal.

\section{The enigma of DQCP and its entanglement solution}
The efforts in computation of 2D DQCP lattice models, both in spin and fermion representations, have by now largely focused on the evaluation of correlation functions related with the order parameters of the symmetry breaking phases on both sides of the DQCP, given them AFM, VBS, QSH and SC orders, and then make comparison with the theoretical predictions or bounds on the critical properties. This turns out to be a very subtle and complicated endeavour, and in fact, the present enigmatic situation of the {AFM-VBS} DQCP, where such comparison often-time brought contradiction and inconsistency, is largely due to the limitation of the present day already sophisticated and cutting edge large-scale numerical computation on finite size lattices, and the yet still-far-away distance of these finite size lattice models from the continuum limit where the field theoretical tools are at working~\cite{senthilQuantum2004,senthilDeconfined2004,qinDuality2017,wangDeconfined2017}. Over the years, it is realized such a dilemma is very vividly presented in DQCP, much more so compared with the situation in conventional QCPs.

Facing the conundrum between the finite sizes and thermodynamic limit, people try to find new probes that could potentially fill in the gap between the lattice model and continuum theory, and the {\it entanglement measurement} stands out as one of the most sensitive choice that could potentially resolve the dilemma and provide decisive answers. Previous successful attempts include the computation of the R\'enyi entanglement entropy (EE) and the disorder operators at the DQCP of JQ model~\cite{zhaoScaling2022,wangScaling2021}, and the latter for the QSH-SC fermion DQCP model~\cite{liuFermion2022}. In all these works, the ubiquitous message from the unbiased numerical simulations is that, the {VBS-AFM DQCP in the JQ model~\cite{zhaoScaling2022,wangScaling2021} and the QSH-SC DQCP in} fermion model~\cite{liuFermion2022} are both {\it not unitary conformal field theory (CFT), or more likely, not a CFT at the first place}. This statement is based on the unanimous observation that the universal coefficients of the logarithmic corrections (related with the central charge of the CFT) in EE and disorder operator exhibit the opposite sign compared with the requirement of a CFT~\cite{Casini2012}.

\begin{figure}[htp!]
	\includegraphics[width=\columnwidth]{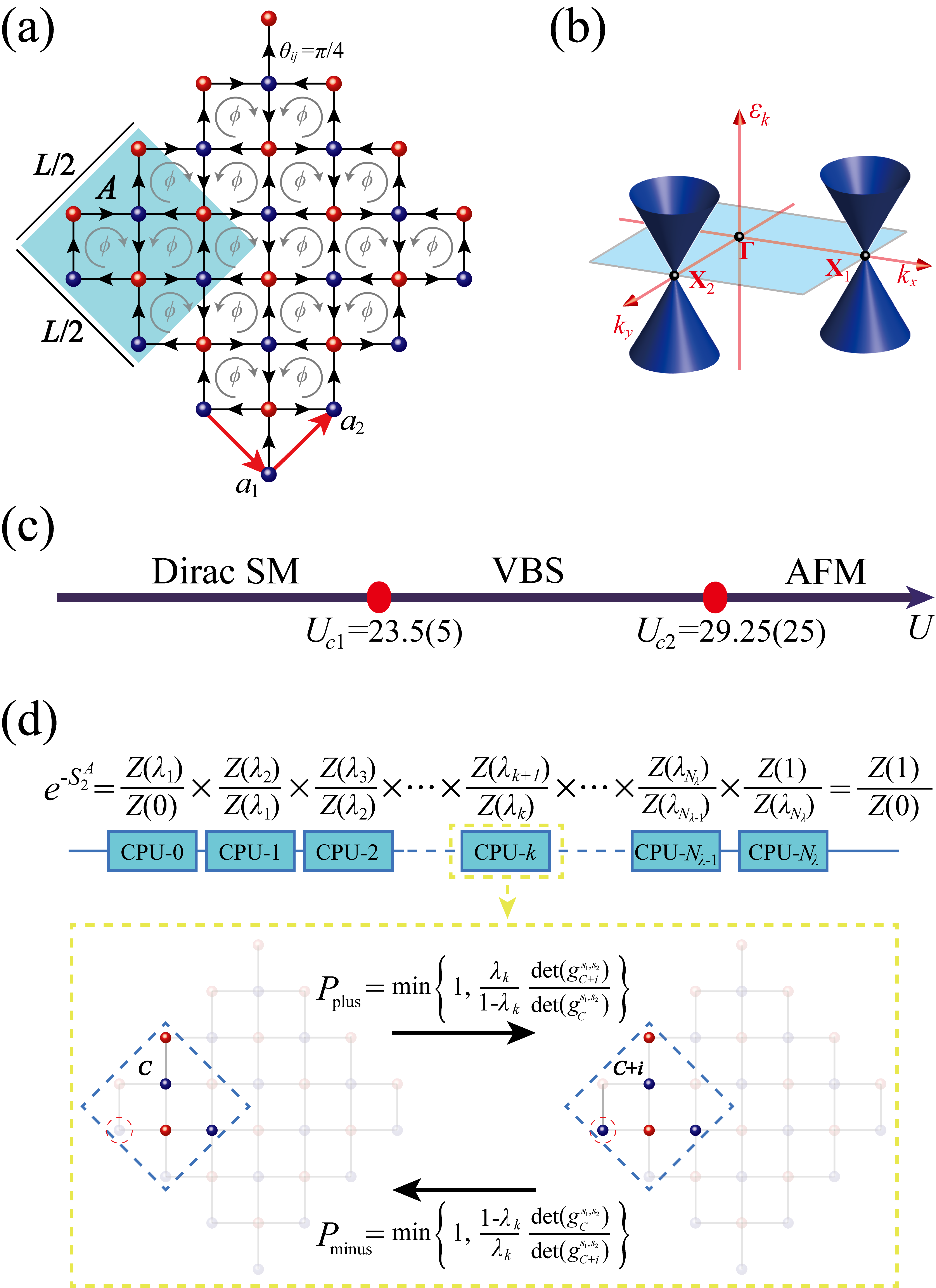}
	\caption{\textbf{(a) The fermion $\pi$-flux plaquette model.} 
		We set the hopping amplitude $t_{ij}=te^{i\theta_{ij}}$ and $\theta_{ij}=\pi/4$ along the black arrows to thread a staggered flux $\phi=\pm\pi$ penetrating each $\square$ plaquette. The unit vectors $a_1$ and $a_2$ are represented with red arrows. The red and blue balls denote the two sublattices. The shaded square is the entanglement region $A$ that we choose to compute the $S_2^{A}$ in a $L\times L$ unit cell lattice with torus geometry.
		\textbf{(b) The Dirac fermions in the first BZ.} 
		The Dirac cones are located at $X_1=(\pi,0)$ and $X_2=(0,\pi)$
		, and $\Gamma=(0,0)$ is also denoted.
		\textbf{(c) The phase diagram of $\pi$-flux plaquette model.} 
		Our previous DQMC simulation has revealed a Gross-Neveu QCP at $U_{c1} = 23.5(5)$ and a {SU(2) VBS-AFM} DQCP at $U_{c2} = 29.25(25)$ ~\cite{liaoDiracI2022}. 
		\textbf{(d) The sketch of incremental algorithm to compute EE.} 
		To obtain accurate value of $e^{-S_2^A}$, we compute each piece of replicated partition function $\frac{Z(\lambda_{k+1})}{Z({\lambda_{k}})}$ independently in parallel according to Eq.~\eqref{eq:eqS8} in Method section, here $\lambda_{k}\in (0,1)$ and $\lambda_{k+1}>\lambda_{k}$. 
		In the incremental process, we update the region $C$ with stochastic process by adding and removing a site into region $C$ with probabilities $P_{\text{plus}}$ and $P_{\text{minus}}$.
		More details of the algorithmic are given in the Method section and the Supplementary Information (SI)~\cite{suppl}.}
	\label{fig:fig1}
\end{figure}  

To better clarify such complicated and somewhat confusing situation, one would need to first develop robust computation scheme of the EE in the lattice model QMC simulations (more for fermionic models, as the robust and high-precision computation of EE for the boson/spin models are easy by now, see below). It is well-known that for interacting fermion models, such a task is very challenging~\cite{groverEntanglement2013,assaadEntanglement2014,assaadStable2015,broeckerNumerical2016,peterRenyi2014}.  This is mainly due to the fact that, the EE, for example the $n$-th order R\'enyi EE $S_n$, needs to be computed in path-integral QMC simulations with a replicated manifold with the $n$ connected copies in the entanglement region. Since the typical DQMC simulation for interacting fermions is already very expensive $(O(\beta N^3))$, the construction of the replicated manifold and the ensemble average therein~\cite{assaadEntanglement2014,assaadStable2015,parisenEntanglement2018}  
is even more so. As a result, small system sizes in 2D and often noisy data are insufficient to extract the logarithmic corrections, which contain the universal CFT information, in EE beyond the non-universal and leading area law contribution.

The situation is improving very rapidly, the algorithmic development in entanglement computation for quantum many-body lattice models at higher dimensions, is simply astonishing~\cite{dEmidioEntanglement2020,zhaoMeasuring2022,zhaoScaling2022,yanWormhole2021,songReversing2022,dEmidioUniversal2022}. It is the development of the nonequilibrium incremental algorithm, which converts the computation of the second R\'enyi EE $S_2$ into the parallel execution of 
Jarzynski equality~\cite{albaOut2017,dEmidioEntanglement2020,Jarzynski1997,zhaoScaling2022} of the free energy difference between the partition functions on the replicated manifolds, makes the precise determination of EE on JQ and other 2D quantum spin models easy and robust, and controlled results with the expected CFT information are obtained~\cite{dEmidioEntanglement2020,zhaoMeasuring2022,zhaoScaling2022}, such that at present the computation of EE in 2D quantum spin systems with good data quality is becoming a standard practice~\cite{zhaoMeasuring2022,zhaoScaling2022}. More recently, similar implementation, based on the Zwanzig's formula for classical free energy difference~\cite{zwanzigHigh1954}, is available for DQMC on fermion lattice model~\cite{dEmidioUniversal2022}. 

In this work, we further develop such computation scheme into its incremental and parallel version and to compute the EE for 2D fermionic $\pi$-flux plaquette model on square lattice~\cite{liaoDiracI2022,liaoDiracII2022,liaoDiracIII2022}, where from conventional order parameter measurements, a DQCP is reported to exist between a VBS and a AFM phase. Our high precision EE measurements reveal the coefficient of the logarithmic correction therein, is again negative, and hence this DQCP, together with its JQ and QSH-SC fermionic model~\cite{liuFermion2022} cousins, is also {\it not a CFT}.  Based on these unanimous results from both fermion and spin systems, we conclude that the available 2D {SU(2) AFM-VBS DQCPs are not conformal fixed point, from the perspective of quantum entanglement.} Our observations naturally explain the experimental difficulties in finding DQCP in the VBS-AFM and QSH-SC systems by now. New proposals~\cite{christosModel2023} and understandings of these seemingly quantum critical yet not conformal transition points, are therefore called for.

\section{Results}
{\noindent {\it Model and Method.}---}
We study the 2D fermionic $\pi$-flux plaquette model at half-filling on square lattice~\cite{liaoDiracI2022,liaoDiracII2022,liaoDiracIII2022,ouyangProjection2021} with the Hamiltonian
\begin{equation}
H=-\sum_{\langle i j\rangle, \sigma} \left( t_{ij}  c_{i \sigma}^{\dagger} c_{j \sigma}+\text { H.c. }\right)+U \sum_{\square}\left(n_{\square}-1\right)^2,
\label{eq:eq1}
\end{equation}
where $t_{ij}=te^{i\theta_{ij}}$ is the nearest-neighbor hopping amplitude and we set $t=1$ as the energy unit, $c_{i \sigma}^{\dagger}$ and $c_{i \sigma}$ represent the creation and annihilation operators for fermions on site $i$ with spin $\sigma=\uparrow,\downarrow$,
$n_\square\equiv \frac{1}{4} \sum_{i \in \square} n_{i}$ denotes the extended particle number operator of $\square$-plaquette with $n_{i}= \sum_{\sigma} c_{i \sigma}^{\dagger} c_{i \sigma}$ and at half-filling we have $\langle n_{\square}\rangle =1$. A single electron will obtain a flux $\phi=\sum_{\square}\theta_{ij}$ penetrating each $\square$ plaquette when it hops around, and we focus on the case $\phi=\pi$ by giving each bond a phase $\pi/4$ as shown Fig.~\ref{fig:fig1} (a), which results in two Dirac cones located at $X_1=(\pi,0)$ and $X_2=(0,\pi)$ in the first Brillouin zone (BZ), as shown in Fig.~\ref{fig:fig1} (b). 
We simulate the systems with linear size $L=4,6,8,10,12,14$ and $16$ and choose the entanglement region $A$ as the shaded square in Fig.~\ref{fig:fig1} (a), which includes $N_A=L^2/2$ sites (note the system has 2 sites per unit cell).

Our previous DQMC simulation of this model has revealed two interesting QCPs as a function of the $U$~\cite{liaoDiracI2022}, as shown in Fig.~\ref{fig:fig1} (c). The first QCP is at $U_{c1} = 23.5(5)$ separating the Dirac semimetal (DSM) phase and a gapped VBS phase with a $Z_{4}$ discrete lattice symmetry breaking.  The transition is of the Gross-Neveu (GN) type and found to have emergent U(1) symmetry as the $Z_4$ anisotropy is irrelevant at the low-energy description of the GN-QCP~\cite{louAntiferromagnetic2009}. The second one is at $U_{c2} = 29.25(25)$, separating two spontaneous symmetry breaking phases, e.g. $Z_4$ for VBS and SU(2) for AFM phases, and is consistent with the properties of DQCP discussed in the introduction~\cite{senthilDeconfined2004,senthilQuantum2004}. One particular interesting point of the DQCP in our model is that the obtained critical exponents $\eta_\text{VBS}=0.6(1)$ and $\eta_\text{AF}=0.58(3)$ actually satisfy the CFT Bootstrap bound~\cite{nakayamNecessary2016,poland2019conformal} for the emergent continuous symmetry. It is based on such knowledge, that in this work, we carry out the large-scale QMC investigation of the EE at both the DSM phase and DQCP ($U_{c2}$).

\begin{figure}[htp!]
\includegraphics[width=\columnwidth]{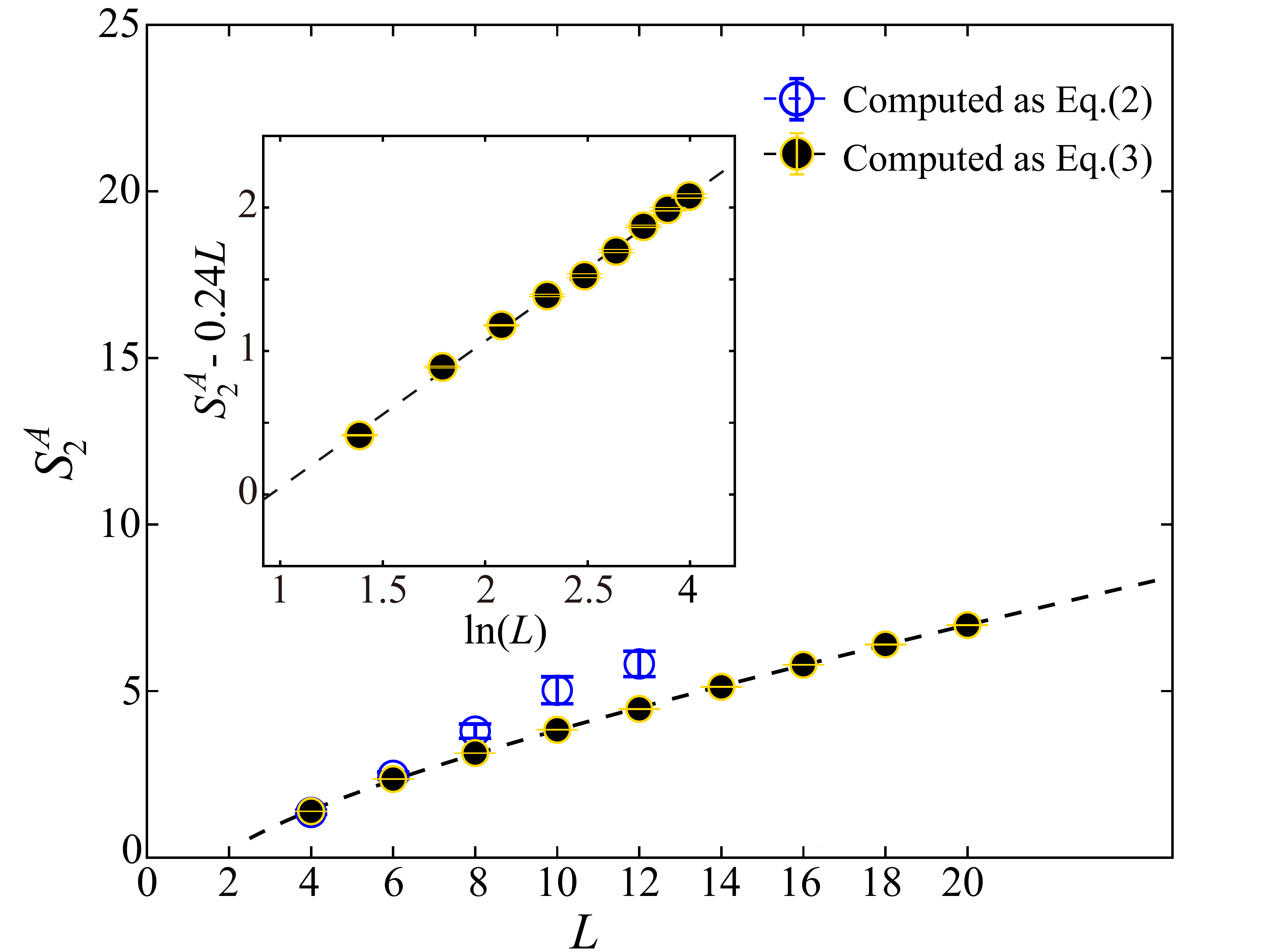}
\caption{\textbf{Benchmark on the Square Hubbard model.} We applied two methods to compute the R\'enyi EE $S_{2}^{A}(L)$ for square Hubbard model at half-filling. Here the entanglement region is half the lattice and $U=8$. The blue circles correspond to original definition in Eq.~\eqref{eq:eq2}, which has larger deviation at larger system sizes, due to the bias in the sampling process. While our incremental algorithm in Eq.~\eqref{eq:eq3} give the correct black dots, as it performs {\it both the correct important sampling and parallel execution}. We fit the black dot data with Eq.~\eqref{eq:eq4} and obtain the coefficient of the logarithmic term $s=-1.0(2)$, which is consistent with the theoretical value $\frac{N_G}{2}=1$, where $N_G=2$ is the number of Goldstone modes in the AFM phase. The black dash line represents the fitting result $S_2^{A}(L)= 0.24(2) L+ 1.0(2)\ln (L)-1.0(3)$. Inset: The EE subtracting off the area law part, $S_2^A - 0.24L$, as function of ln$(L)$.}
	\label{fig:fig2}
\end{figure}  

\begin{figure*}[htp!]
	\includegraphics[width=2\columnwidth]{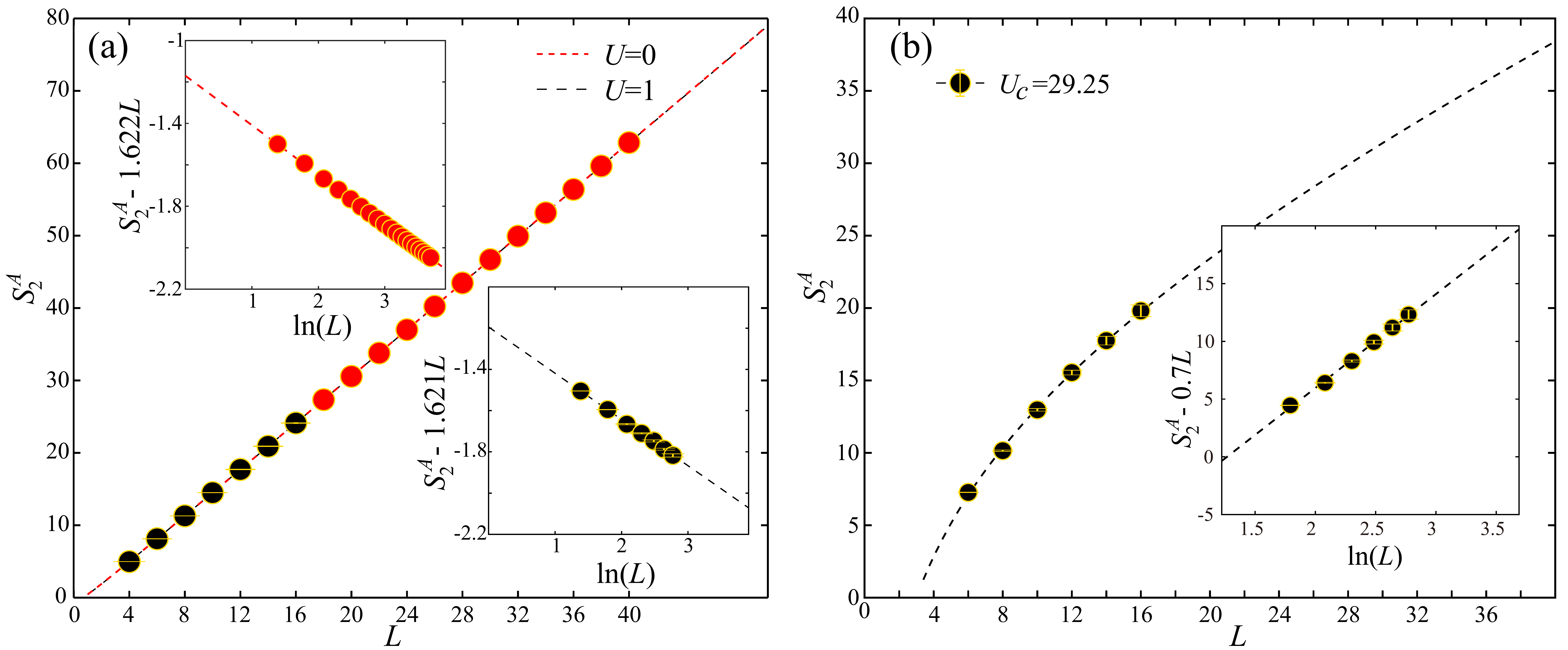}
	\caption{\textbf{R\'enyi entanglement entropy $S_{2}^{A}$ for the Dirac CFT and the non-unitary DQCP}. 
(a) The main panel shows results obtained from our incremental algorithm at $U=0$ ($U=1$) up to system size $L=40$ ($L=16$).
The fitting result according to Eq.~\eqref{eq:eq4} are $S_2^A(L) = 1.622(1)L-0.239(1)\ln(L)-1.170(4)$ for the $U=0$ case, and $S_2^A(L) = 1.621(5)L-0.22(3)\ln(L)-1.19(5)$ for the $U=1$ case.
We found the universal corner coefficients $s$, for both $U=0$ and $U=1$ cases, are all well consistent with the theoretical result of free Dirac CFT with $s\approx 0.23936$~\cite{helmesUniversal2016}.
Left inset shows $S_2^A - 1.622L$, the EE subtracting off the area law part, as function of ln$(L)$ for the $U=0$ case, while right inset shows $S_2^A - 1.621L$ as function of ln$(L)$ for the $U=1$ case.  
(b) The main panel shows $S^{A}_2(L)$ at the $U_{c2}=29.25$ DQCP. The fitting result at $U_{c2}$ is $S_2^{A}(L)= 0.7(3) L+ 6(3)\ln (L)-7(4)$. Inset shows the $S_2^{A}-0.7 L$ versus $\ln (L)$ such that the sign of the log-corrections manifest, $s<0$ according to Eq.~\eqref{eq:eq4}, which means that {this SU(2) VBS-AFM} DQCP is not an unitary CFT as it is in contrast with the positivity requirement of EE~\cite{Casini2012}. }
	\label{fig:fig3}
\end{figure*}  

The reliable computation of the EE cannot be achieved without the developments of the incremental algorithm~\cite{dEmidioEntanglement2020,zhaoScaling2022,zhaoMeasuring2022,dEmidioUniversal2022}, as schematically shown in Fig.~\ref{fig:fig1} (d). In this algorithm, we first introduce two independent DQMC auxiliary field configurations $\{s_1\}$ and $\{s_2\}$ and compute the 2nd R\'enyi EE~\cite{groverEntanglement2013} as,
\begin{equation}
e^{-S_2^A}=\sum_{\left\{s_1\right\},\left\{s_2\right\}} P_{s_1} P_{s_2} \operatorname{det} g_A^{s_1,s_2},
\label{eq:eq2}
\end{equation}
where $P_s$ is the normalized weight of configuration $\{s\}$, $A$ is the $\frac{L}{2}\times \frac{L}{2}$ entanglement region in Fig.~\ref{fig:fig1} (a), $g_A^{s_1,s_2}= G_A^{s_1} G_A^{s_2}+\left(\mathbb{I}-G_A^{s_1}\right)\left(\mathbb{I}-G_A^{s_2}\right) $ is referred as the Grover matrix in Ref.~\cite{dEmidioUniversal2022}. We then parametrize the evaluation of the Eq.~\eqref{eq:eq2} into an incremental process of $\lambda$ evolving from 0 to 1, corresponding to the subset $C$ of the entanglement region growing from $\varnothing$ to $A$, as
\begin{equation}
\begin{aligned}
e^{-S_2^A}&=\frac{\sum_{\left\{s_1\right\},\left\{s_2\right\}} W_{s_1} W_{s_2} \operatorname{det} g_{C=A}^{s_1,s_2}}{\sum_{\left\{s_1\right\},\left\{s_2\right\}} W_{s_1} W_{s_2} \operatorname{det} g_{C=\varnothing}^{s _1,s_2}}\\
&=\frac{Z(\lambda_1)}{Z(0)}\frac{Z(\lambda_2)}{Z(\lambda_1)}\cdots\frac{Z(\lambda_{k+1})}{Z(\lambda_k)} \cdots\frac{Z(1)}{Z(\lambda_{N_\lambda})},
\end{aligned}
\label{eq:eq3}
\end{equation}
in which $Z(\lambda=0)$ is the partition function of the two disconnected replicas (with the subset $C=\varnothing$) and $Z(\lambda=1)$ is the replicated partition function where the entanglement region $A$ is fully connected in imaginary time direction (with $C=A$). And $W_s$ is the unnormalized weight. As in our computation of EE in spin/boson systems in Refs.~\cite{zhaoScaling2022,zhaoMeasuring2022}, each piece in the product of Eq.~\eqref{eq:eq3} can be computed independently in parallel, as shown in Fig.~\ref{fig:fig1} (d), and inside each piece, we increase or decrease the entanglement region in a stochastic process as suggested in Ref.~\cite{dEmidioUniversal2022}. 
We note in the original method of entanglement computation in Eq.~\eqref{eq:eq2}, the weight is the joint distribution of the auxiliary field $s_1$ and its replica $s_2$, where the measurement is the determinant of the Grover matrix: $e^{-S_2^A}=\langle \operatorname{det} g_A^{s_1,s_2}\rangle$. Since the Grover matrix can be either singular or with broad distribution, it is difficult to accurately sample the values which are always exponentially small. While in the incremental algorithm in Eq.~\eqref{eq:eq3}, one writes the determinant into the weight, and introduce $\lambda$ to parameterize the parallel pieces of the partition function ratio, so that the measurement of each piece is independent and gives rise to a controllable value at $O(1)$. Moreover, as discussed in SI~\cite{suppl}, an elegantly designed update scheme reduces computational complexity from $O(\beta N^4)$ to  $O(\beta N^3)$.

It is the correct important sampling of incremental processes within each parallel piece and their independent executions, in a "divide-and-conquer" strategy, greatly suppressed the noise and overcome the original difficulties both in the data quality and computational complexity of EE. With the incremental algorithm, one can finally compute the EE in interacting fermion systems within the DQMC framework~\cite{dEmidioUniversal2022} and extract the universal scaling parts of the 2D entanglement region beyond the area law contribution. Detailed descriptions of our algorithm are given in the Method section, SI~\cite{suppl} and Ref.~\cite{panComputing2023}.

\vspace{0.2cm}
{\noindent {\it Benchmark on the Square Hubbard model.}---To offer a benchmark before showing the main results, in Fig.~\ref{fig:fig2}, we compare the R\'enyi EE of the half-filled square Hubbard model with $U=8$ calculated by these two methods. Here the entanglement region is half of the lattice: $L/2 \times L$. One clearly sees that when the system size $L$ is small, the results of the two methods coincide, but when the size gradually increases, the data errorbar of the former becomes larger and  the values also have obvious bias.
	
Theoretical analysis can help one to appreciate the correctness of the incremental algorithm. Since the half-filled square lattice Hubbard is always in the AFM phase ($U>0$), the R\'enyi EE of the system with spontaneous broken SU(2) continuous symmetry should have a form in Eq.~\eqref{eq:eq4} with the universal log-coefficient $s=-\frac{N_G}{2}$ ~\cite{metlitskiEntanglement2011,laflorencieQuantum2016}, where $N_G=2$ is the number of the Goldstone modes. As shown in Fig.~\ref{fig:fig2}, the coefficient before the logarithmic term obtained by the second method is $1.0(2)$, and well consistent with the theoretical expected value $1$.

\vspace{0.2cm}
{\noindent {\it Results on DSM.}---}
We start from the EE at the DSM limit ($U=0,1$) of the model in Eq.~\eqref{eq:eq1}. Fig.~\ref{fig:fig3} (a) shows the obtained $S^A_2$ as
\begin{equation}
S_2^{A}(L)=a L-s \ln (L)-b,
\label{eq:eq4}
\end{equation}
where $a$ is the coefficient of the area law term, $s$ is the universal constant that contains the CFT information and the geometric partition between the entanglement region $A$ and the environment $\overline{A}$~\cite{fradkinEntanglement2006,cardyFinite1988,calabreseEntanglement2004}, and $b$ is a nonuniversal constant. For the free Dirac fermions, one 90$^{\circ}$ corner of $A$ contributes to an universal coefficient of 0.01496 to the $\log$ term in the Dirac CFT~\cite{helmesUniversal2016}. Since in our model, rectangle region  $A$ has four 90$^{\circ}$ corners and we have two spin flavors and two Dirac cones in the BZ. The final contribution to the $\log$ coefficient is $s_0 = 16 \times 0.01496 \approx 0.23936$. Our computed results in Fig.~\ref{fig:fig3} for $U=0$ case fully agree with the theoretical expectation. In the main panel, the $S^A_2(L)$ with $L$ upto 40 clearly exhibits a dominate area law behavior (linear in $L$), and by fitting the data of $S^A_2 - a L$ vs $\ln(L)$ as shown in the left inset of Fig.~\ref{fig:fig3} (a), a converged slope of $s=0.239(1)$ is clearly obtained. 
We also computed the $S^A_2$ at $U=1$ inside the DSM phase, although with a finite plaquette interaction, the system still flows to the free Dirac CFT in the low-energy limit, and it is indeed the case that, our $S^A_2$ results in Fig.~\ref{fig:fig3}(a) (the right inset by fitting the $S^A_2-aL$ vs $\ln(L)$) also give rise to a precise log-coefficient of $s=0.22(3)$.

\vspace{0.2cm}

{\noindent {\it {Results on the VBS-AFM DQCP.}}---}
Fig.~\ref{fig:fig3} (b) is the $S^A_2$ at the {SU(2) VBS-AFM} DQCP. One can already see from the main panel that the $S^A_2(L)$ exhibits different scaling  behavior compared with that in Fig.~\ref{fig:fig3} (a), namely, there exists a nonnegligible logarithmic corrections such that the $S^A_2(L)$ has a small downwards bending. Such bending has also been observed in the $S^A_2(L)$ of {SU(2) JQ model at its VBS-AFM DQCP}~\cite{zhaoScaling2022} and it is the signature that the log-correction has anomalous coefficient. In the insets of Fig.~\ref{fig:fig3} (b), we also subtract the area law contribution and plot $S^A_2 - aL$ vs $\ln(L)$, now we clearly see that the slope of the curve in the inset has a obvious negative sign compared with that in the inset of Fig.~\ref{fig:fig3} (a). 
And not only the sign is reversed, we find that the value of the negative log-coefficient here $s=-6$ is more than 20 times larger than that of the free Dirac $s\approx 0.23936$. Such large and negative log-coefficient is at odds with the positivity requirement of the entanglement entropy for a CFT~\cite{Casini2012} and offers a clear signature that the DQCP here is non-unitary, with exactly the same behavior as have been observed in the JQ model~\cite{zhaoScaling2022} and the other fermion DQCP model~\cite{liuFermion2022}. These results, combining the observation of the negative log-correction in EE and disorder operators in both spin and fermion DQCP models~\cite{wangScaling2021,zhaoScaling2022,liuFermion2022}, really deliver a clear message that, all the 2D {SU(2) AFM-VBS DQCP lattice models} we have so far, cannot pass the scrutiny of quantum entanglement measurements and therefore they are not of a CFT description.

\vspace{0.2cm}
\noindent{{Discussions}} Since our observation of a large and negative $s$ is not allowed in an unitary CFT, it in fact explains the recent tension between the conformal bootstrap bounds following from unitary conformal invariance and the numerically computed critical exponents in {2D SU(2) VBS-AFM} DQCP models~\cite{shaoQuantum2016,nahumDeconfined2015,nahumNote2020}. Although it is natural to connect the observed regime of the DQCP with a non-unitary fixed point very close to the physical parameter space~\cite{maTheory2020,nahumNote2020,wangPhases2021}, which implies approximate conformal invariance within a large length scale and the "non-unitarity" manifesting in the imaginary part of scaling dimensions should be quite small. However, it is not clear whether such a scenario can naturally explain the large and negative $s$ observed here (more than 20 times larger than the positive $s$ in the closeby Dirac CFT) and also in the JQ model~\cite{zhaoScaling2022}. In the known example of weakly first-order transition controlled by a complex CFT (note the suggestion that DQCP is the precursors to a weakly first-order transition~\cite{jiangFrom2008,kuklovDeconfined2008,chenDeconfined2013,dEmidioDiagnosing2021}), such as the $Q=5$ Potts model in (1+1)d, the log-coefficient of EE is a small and real positive number~\cite{maShadow2019}. Our consistent results however suggest that the violation of unitarity in JQ and fermionic DQCP models is not just a small complex correction.

On the experimental front, the efforts by now do not find signature in the form of {2D VBS-AFM and QSH-SC DQCPs} beyond the Landau-Ginzburg-Wilson paradigm. For example, following the VBS-AFM DQCP in the spin JQ model,  the pressure driven thermodynamic and dynamic measurements in SrCu$_2$(BO$_3$)$_2$ and similar magnetic compounds~\cite{zayed4spin2017,guoQuantum2020, jimenezquantum2021,cuiProximate2022}, have mostly found first order transition. And the experimental realization of the QSH-SC fermion DQCP, such as unconventional superconducting quantum criticality in 2D material, is actually more consistent with an intermediate phase between the QSH and SC, instead of a DQCP. Our results naturally explain the experimental difficulties as both spin and fermion models of DQCP are not DQCP in the first place. One shall look for other materials which can pass the entanglement examination. {We note, however, the recent proposal of a conformal 2D DQCP with SO(5)$_f \times$ SO(5)$_b$ global symmetry that describes the cuprate phase diagram with pseudogap metal, d-wave superconductivity and charge order as various symmetry breaking phases~\cite{christosModel2023}, is certainly of great importance to investigate its validity.} 

Our observations post a fundamental question, that, all the available {VBS-AFM and QSH-SC} lattice models in 2D which were found to acquire a DQCP both in spin representation such as the JQ model~\cite{zhaoScaling2022,wangScaling2021}, or in fermion representation such as the QSH-SC model without the monople and two-length scale subtleties~\cite{liuFermion2022} or the present $\pi$-flux plaquette VBS-AFM model seeming satisfying the bootstrap bound for emergent continuous symmetry, cannot pass the examination from the entanglement perspective, and therefore, unfortunately, none of them is an unitary CFT. Such conclusions on the one hand show how sensitive the entanglement tools are in detecting the fundamental properties of exotic and complicated lattice models, and on the other are calling for new and ultimate understanding of these 2D DQCP lattice models {such as the new proposal in Ref.~\cite{christosModel2023}}. However enigmatic the situation might seems with our discoveries, the unanimous observation of the non CFT nature of these 2D DQCP models, marks the end of the beginning of unambiguous finding of the quantum phase transitions truly beyond the paradigm of Landau-Ginzburg-Wilson, even since its suggestion two decades ago~\cite{senthilQuantum2004,senthilDeconfined2004}. 

\section{Method}

\noindent{\it projector DQMC simulation.} 
In this work, we focus on calculating the 2nd order R\'enyi entanglement entropy $S_2^A$ of a $\pi$-flux plaquette model on 2D square lattice with $N=2L^2$ sites. 
Since $S_2^A$ is a ground-state observable, it is very suitable to empoly the projector DQMC method to compute the quantity.

In this method, we obtain the ground state $\vert \Psi_0 \rangle$ with the help of the projection of a trial wave function $\vert \Psi_T \rangle$, i.e. $\vert \Psi_0 \rangle = \lim\limits_{\Theta \to \infty} e^{-\Theta H} \vert \Psi_T \rangle$. The scheme is very flexible and the only requirement is that $\vert \Psi_T \rangle$ must be nonorthogonal to $\vert \Psi_0 \rangle$.

Physical observable $\hat{O}$ can be calculated as
\begin{equation}
\label{eq:eq5}
\langle \hat{O} \rangle = \frac{\langle \Psi_0 \vert \hat{O} \vert \Psi_0 \rangle}{\langle \Psi_0 \vert \Psi_0 \rangle} 
= \lim\limits_{\Theta \to \infty} \frac{\langle \Psi_T \vert  e^{-\Theta H } \hat{O}  e^{-\Theta H} \vert \Psi_T \rangle}{\langle \Psi_T \vert  e^{-2\Theta H} \vert \Psi_T \rangle} .
\end{equation}
where $H$ is the Hamiltonian and $2\Theta$ is the projection length.

Since $H$ consists of the non-interacting and interacting parts, $H_0=-\sum_{\langle i j\rangle, \sigma} \left( t_{ij}  c_{i \sigma}^{\dagger} c_{j \sigma}+\text { H.c. }\right)$ and $H_U=U \sum_{\square}\left(n_{\square}-1\right)^2$, respectively, that do not commute. We perform Trotter decomposition to discretize the projection length $2\Theta$ into $L_\tau$ imaginary time slices ($2\Theta=L_\tau \Delta_\tau$) and have
\begin{equation}
\langle\Psi_{T}|e^{-2\Theta H}|\Psi_{T}\rangle=\langle\Psi_{T}|\left(e^{-\Delta_\tau H_{0}}e^{-\Delta_\tau H_{U}}\right)^{L_\tau}|\Psi_{T}\rangle+\mathcal{O}(\Delta{\tau}^{2}),
\label{eq:eq6}
\end{equation}
where the non-interacting part $H_0$ and interacting part $H_U$ are separated. 
Since the Trotter decomposition process will give rise to a small systematic error $\mathcal{O}(\Delta_{\tau}^2)$, we need to set $\Delta_\tau$ as a small number to get accurate results.
$H_U$ contains the quartic fermionic interactions that we employ a SU(2) symmetric Hubbard-Stratonovich (HS) decomposition to decouple it with the auxiliary fields $\{s_{\square,l_\tau}\}$ to a fermion bilinear, in our case, the charge density operator.

In our implementation, the HS decomposition at time slice $l_\tau$ is described by
\begin{equation}
e^{-\Delta_\tau U(n_{\square}-1)^{2}}=\frac{1}{4}\sum_{\{s_{\square,l_\tau}\}}\gamma(s_{\square,l_\tau})e^{\alpha\eta(s_{\square,l_\tau})\left(n_{\square}-1\right)}
\label{eq:decompo}
\end{equation}
with $\alpha=\sqrt{-\Delta\tau U}$, $\gamma(\pm1)=1+\sqrt{6}/3$,
$\gamma(\pm2)=1-\sqrt{6}/3$, $\eta(\pm1)=\pm\sqrt{2(3-\sqrt{6})}$,
$\eta(\pm2)=\pm\sqrt{2(3+\sqrt{6})}$, and the sum symbol is taken over the auxiliary field $s_{\square,l_\tau}$ on each square plaquette at time $l_\tau$. Then, we have
\begin{widetext}
   \begin{eqnarray}
\langle\Psi_{T}|e^{-2\Theta H}|\Psi_{T}\rangle=\sum_{\{s_{\square,l_\tau}\}}\left[\left(\prod_{l_\tau}^{L_{\tau}}\prod_{\square}^{N}\gamma(s_{\square,l_\tau})e^{\alpha\eta(-s_{\square,l_\tau})}\right)\det\left[P^{\dagger}B^{s}(2\Theta,0)P\right]\right]
\label{eq:mcweight}
   \end{eqnarray}
\end{widetext}
where $P$ is the coefficient matrix of trial wave function $|\Psi_T\rangle$; $B^{s}(2\Theta,0)$ is defined as
\begin{equation}
B^{s}(\tau_{2},\tau_{1})= \prod_{l_\tau=l_{1}+1}^{l_{2}} \left( \mathrm{e}^{-\Delta_\tau H_0} \prod_{\square}^{N} \mathrm{e}^{\alpha\eta(s_{\square,l_\tau}) n_{\square}} \right)
\end{equation}
with $l_1 \Delta_\tau = \tau_1$ and $l_2 \Delta_\tau = \tau_2$, and has a property $B^{s}(\tau_3,\tau_1)=B^{s}(\tau_3,\tau_2)B^{s}(\tau_2,\tau_1)$.

We further introduce the notation $B^{\rangle}_\tau = B^{s}(\tau,0)P$ and $B^{\langle}_\tau =P^\dagger B^{s}(2\Theta,\tau)$.
The equal time Green's functions are given as
\begin{equation}
G^{s}(\tau)=\mathbb{I}-B^{\rangle}_\tau \left(B^{\langle}_\tau B^{\rangle}_\tau \right)^{-1}B^{\langle}_\tau,
\end{equation}
where $\mathbb{I}$ is the identity matrix.
On finite precision computer, because of the repeated multiplications between matrices with exponentially large and small values, a straightforward calculations of $B^{\rangle}_\tau$  and $B^{\langle}_\tau$  would lead to serious numerical instabilities.
To circumvent this problem, the UDV matrix decompositions~\cite{assaadWorld-line2008,lohStable1992} are introduced in the projector DQMC method.
In practice, we decompose $B^{\rangle}_{\tau}$ into $B^{\rangle}_{\tau} = U^{\rangle}_{\tau} D^R_{\tau} V^R_{\tau}$ and, similarly, decompose $B^{\langle}_\tau$ into $B^{\langle}_\tau =V^L_{\tau} D^L_{\tau} U^{\langle}_\tau $. 
Then the equal time Green's functions are given more accurately
\begin{equation}\begin{aligned}
G^{s}(\tau)&=\mathbb{I}-B^{\rangle}_\tau \left(B^{\langle}_\tau B^{\rangle}_\tau \right)^{-1}B^{\langle}_\tau \\
 &= \mathbb{I}-U^{\rangle}_\tau \left(U^{\langle}_\tau U^{\rangle}_\tau \right)^{-1}U^{\langle}_\tau,
\label{eq:ns}
\end{aligned}\end{equation}
because the $D$ and $V$ matrices, which contain the extremely large and small eigenvalues, are cancelled after every numerical stabilization procedure.

The imaginary-time displaced Green's functions could be calculated in projector DQMC method as
\begin{equation}
\begin{aligned}
&\left\{\begin{array}{l}
G^s\left(\tau_2, \tau_1\right)=B^s\left(\tau_2, \tau_1\right) G^s\left(\tau_1\right), \\
G^s\left(\tau_1, \tau_2\right)=-\left[\mathbb{I}-G^s\left(\tau_1\right)\right] B^s\left(\tau_2, \tau_1\right)^{-1}, \\
\end{array}\right.  \ \ \ \  \tau_2>\tau_1, \\
&\left\{\begin{array}{l}
G^s\left(\tau_2, \tau_1\right)=- B^s\left(\tau_1, \tau_2\right)^{-1} \left[\mathbb{I}-G^s\left(\tau_1\right)\right], \\
G^s\left(\tau_1, \tau_2\right)=G^s\left(\tau_1\right) B^s\left(\tau_1, \tau_2\right) , \\
\end{array}\right. \ \ \ \  \tau_2<\tau_1. \\
\end{aligned}
\end{equation}
For example, for $\tau_2 > \tau_1$ case, the imaginary-time displaced Green's functions $G^s\left(\tau_2, \tau_1\right)$ will also suffer the problem of numerical instabilities, if one obtains $G^s\left(\tau_2, \tau_1\right)$ through a direct multiplication of $G^s\left(\tau_1\right)$ with $B^s\left(\tau_2, \tau_1\right)$ when $\tau_2$ is much larger than $\tau_1$.
To circumvent this problem, an efficient and easy-to-implement numerically stable method has been developed~\cite{feldbacherEfficient2001} in projector DQMC.
This method is based on the simple observation that equal time Green's function is a projector, $G^s\left(\tau\right) \cdot G^s\left(\tau\right) = G^s\left(\tau\right)$, and can give rise to numerically well defined imaginary-time displaced Green's functions as
\begin{equation}
G^{s}\left(\tau_2,\tau_1 \right)= \prod_{n=0}^{N_d}  G^s\left( \tau^\prime+[n+1]\tau^0, \tau_1+n\tau^0 \right),
\end{equation}
where $\tau^0$ is a small number, $\tau_2=\tau^\prime+[N_d+1]\tau^0$.

In practice, we choose the ground state wavefunction of $H_0$ with infinitesimal flux threaded through the torus as the trial wave function. 
The measurements are performed near $\tau=\Theta$, we set projection time $2\Theta = L$, discrete time slice $\Delta_\tau=0.1$ and $\tau^0=10\Delta_\tau$. More technique details of the projector DQMC method and its implementation of the fermion $\pi$-flux plaquette model, please refer to Refs.~\cite{assaadWorld-line2008,liaoDiracI2022}. 

\vspace{0.2cm}
\noindent{\it Incremental algorithm for EE computation in interacting fermions.} Grover's pioneering work~\cite{groverEntanglement2013} addressed a general method by introducing two independent replica configurations of auxiliary field to calculate the 2nd R\'enyi EE for an interacting fermionic system within the DQMC framework, as shown in Eq.~\eqref{eq:eq2}.
To obtain the 2nd R\'enyi EE with higher precision in DQCP, D’Emidio et al. developed an improved algorithm recently~\cite{dEmidioUniversal2022},  based on the incremental algorithms of the EE computation in boson/spin systems~\cite{dEmidioEntanglement2020,zhaoScaling2022,zhaoMeasuring2022}.
The method converts the entanglement region fluctuations into a stochastic process and could give rise to the 2nd R\'enyi EE with much better precision in 2D strong correlated fermionic systems, compared with the original proposals in Refs.~\cite{groverEntanglement2013,assaadStable2015,assaadEntanglement2014}. Here we further optimize and develop the incremental algorithm to its parallel version. The brief outline is given below with the full detailed workflow provided in SI~\cite{suppl}. 

The replicated partition function $Z_C$ plays the key role in the EE compuation,
\begin{equation}
Z_C=\sum_{\left\{s_1\right\},\left\{s_2\right\}} W_{s_1} W_{s_2} \operatorname{det} g_C^{s_1, s_2},
\end{equation}
where $ W_{s}\propto \operatorname{det}\left[ P^{\dagger} B^s(2 \Theta, 0) P\right] $ is the standard projector DQMC unnormalized weight of configuration $s$, and there is a simple normalized relation $P_{s} = W_{s} / \sum_{\left\{s\right\}}W_{s} $; $C$ is the subset geometry of the entanglement region $A$.
In projector DQMC, we usually  measure physical observables near $\tau=\Theta$, and here we also define strictly the Grover matrix at time slice $\tau=\Theta$

We further define 
\begin{equation}
Z(\lambda)=\sum_{C \subseteq A} \lambda^{N_C}(1-\lambda)^{N_A-N_C} Z_C
\end{equation}
where $\lambda \in [0,1]$, $N_C$ is the number of sites in $C$, and $Z(1) = Z_{C=A}$ and $Z(0) = Z_{C=\varnothing}$. We have $g_{C=\varnothing}^{s_1,s_2} = \mathbb{I}$, then the 2nd R\'enyi EE has a new form as the ratio of replicated partition functions
\begin{equation}
\begin{aligned}
e^{-S_2^A}&=\sum_{\left\{s_1\right\},\left\{s_2\right\}} P_{s_1} P_{s_2} \operatorname{det} g_A^{s_1,s_2} \\
&=\frac{\sum_{\left\{s_1\right\},\left\{s_2\right\}} W_{s_1} W_{s_2} \operatorname{det} g_{C=A}^{s_1,s_2}}{\sum_{\left\{s_1\right\},\left\{s_2\right\}} W_{s_1} W_{s_2} \operatorname{det} g_{C=\varnothing}^{s_1,s_2}}\\
&=\frac{Z(1)}{Z(0)}\\
&=\frac{Z(\lambda_1)}{Z(0)}\frac{Z(\lambda_2)}{Z(\lambda_1)}\cdots\frac{Z(\lambda_{k+1})}{Z(\lambda_k)} \cdots\frac{Z(1)}{Z(\lambda_{N_\lambda})}.
\end{aligned}
\label{eq:eqS8}
\end{equation}
And each small piece in the RHS of Eq.~\eqref{eq:eqS8} is independently computed as
\begin{small}
\begin{equation}
\begin{aligned}
&\frac{Z(\lambda_{k+1})}{Z(\lambda_{k})}=\frac{\sum_{C \subseteq A} \lambda_{k+1}^{N_C}(1-\lambda_{k+1})^{N_A-N_C} Z_C}{\sum_{C \subseteq A} \lambda_{k}^{N_C}(1-\lambda_{k})^{N_A-N_C} Z_C}\\
&=\frac{\sum_{C \subseteq A} \left(\frac{\lambda_{k+1}}{\lambda_{k}}\right)^{N_C} \left(\frac{1-\lambda_{k+1}}{1-\lambda_{k}}\right)^{N_A-N_C} \lambda_{k}^{N_C}\left(1-\lambda_{k}\right)^{N_A-N_C} Z_C}{\sum_{C \subseteq A} \lambda_{k}^{N_C}\left(1-\lambda_{k}\right)^{N_A-N_C} Z_C}\\
&= \langle \left(\frac{\lambda_{k+1}}{\lambda_{k}}\right)^{N_C}\left(\frac{1-\lambda_{k+1}}{1-\lambda_{k}}\right)^{N_A-N_C} \rangle_{\lambda_{k}}.
\label{eq:eqM18}
\end{aligned}
\end{equation}
\end{small}
Additionally, to obtain more available data, one could also calculate $\frac{Z(\lambda_{k-1})}{Z(\lambda_{k})}= \langle \left(\frac{\lambda_{k-1}}{\lambda_{k}}\right)^{N_C}\left(\frac{1-\lambda_{k-1}}{1-\lambda_{k}}\right)^{N_A-N_C} \rangle_{\lambda_{k}} $ in the same Markov chain for $\lambda_{k}$, which will give extra estimate about  $\frac{Z(\lambda_{k})}{Z(\lambda_{k-1})}$.
With the stochastic process of adding and removing a site $i$ in $C$ with the probabilities $P_{\text{plus}}=\min\{1,\frac{\lambda_k}{1-\lambda_k}\frac{\det(g^{s_1,s_2}_{C+i})}{\det(g^{s_1,s_2}_{C})}\}$ and $P_{\text{minus}}=\min\{1,\frac{1-\lambda_k}{\lambda_k}\frac{\det(g^{s_1,s_2}_{C})}{\det(g^{s_1,s_2}_{C+i})}\}$  shown in Fig.~\ref{fig:fig1} (d). The final product is obtained with parallel simulation $N_{\lambda}$ time faster than the sequential execution of Eq.~\eqref{eq:eqS8}, by varying $\lambda$ from 0 to 1 in $N_{\lambda}$ independent Monte Carlo processes.

As shown in our previous works~\cite{zhaoScaling2022,zhaoMeasuring2022}, such "divide-and-conquer" strategy in distributing the evolution of different pieces of the EE and summation afterwards, greatly suppressed the noise and finally enable one to access the controlled universal scaling analysis of the 2D entanglement region beyond the leading area law contribution. In this work, we further improve the method and carry out the simulation at the DQCP of our model in Eq.~\eqref{eq:eq1}. 
The detailed procedures of algorithmic implementation are given in SI~\cite{suppl}.

\vspace{0.2cm}
\section{Acknowledgements}
We thank Jiarui Zhao, Zheng Yan, Yan-Cheng Wang, Jonathan D'Emidio, Junchen Rong, Meng Cheng, Kai Sun and Fakher Assaad for inspirational discussions and collaborations on the previous related projects. Y.D.L. acknowledges support from National Natural Science Foundation of China (Grant No. 12247114) and the China Postdoctoral Science Foundation (Grants Nos. 2021M700857 and  2021TQ0076).
Y.Q. acknowledges support from the National Natural Science Foundation of China (Grant Nos. 11874115 and 12174068).
G.P.P., W.L.J. and Z.Y.M. acknowledge the support from the Research Grants Council of Hong Kong SAR of China (Grant Nos. 17301420, 17301721, AoE/P-701/20, 17309822), the ANR/RGC Joint Research Scheme sponsored by Research Grants Council of Hong Kong SAR of China and French National Reserach Agency(Porject No. A\_HKU703/22), the Strategic Priority Research Program of the Chinese Academy of Sciences (Grant No. XDB33000000), the K. C. Wong Education Foundation (Grant No. GJTD-2020-01) and the Seed Funding “Quantum-Inspired explainable-AI” at the HKU-TCL Joint Research Centre for Artificial Intelligence.
The authors also acknowledge Beijng PARATERA Tech Co.,Ltd., the HPC2021 system under the Information Technology Services and the Blackbody HPC system at
the Department of Physics, University of Hong Kong for providing HPC resources that have contributed to the research results reported in this paper.

\bibliography{ref.bib}
\bibliographystyle{apsrev4-2}

\clearpage
\onecolumngrid

\begin{center}
\textbf{\large Supplementary Material for ``The teaching from entanglement: 2D SU(2) antiferromagnet to valence bond solid deconfined quantum critical points are not conformal''}
\end{center}
%\appendix
\setcounter{equation}{0}
\setcounter{figure}{0}
\setcounter{table}{0}
\setcounter{page}{1}
\makeatletter
\renewcommand{\theequation}{S\arabic{equation}}
\renewcommand{\thefigure}{S\arabic{figure}}
\renewcommand{\bibnumfmt}[1]{[S#1]}
\renewcommand{\citenumfont}[1]{S#1}
\setcounter{secnumdepth}{3}

In this Supplementary Material, we explain the workflow of the incremental algorithm in computing the EE in projector QMC (PQMC) framework. 

As shown in Eq.~\eqref{eq:eq3} in the main text, we parametrize the evaluation of the $e^{-S^A_2}$ into an incremental process of $\lambda$ evolving from 0 to 1, corresponding to the subset $C$ of the entanglement region growing from $\varnothing$ to $A$, as
\begin{equation}
\begin{aligned}
e^{-S_2^A}&=\frac{\sum_{\left\{s_1\right\},\left\{s_2\right\}} P_{s_1} P_{s_2} \operatorname{det} g_{C=A}^{s_1,s_2}}{\sum_{\left\{s_1\right\},\left\{s_2\right\}} P_{s_1} P_{s_2} \operatorname{det} g_{C=\varnothing}^{s _1,s_2}}\\
&=\frac{Z(\lambda_1)}{Z(0)}\frac{Z(\lambda_k)}{Z(\lambda_1)}\cdots\frac{Z(\lambda_{k+1})}{Z(\lambda_k)} \cdots\frac{Z(1)}{Z(\lambda_{N_\lambda})},
\end{aligned}
\end{equation}
in which $Z(\lambda=0)$ is the partition function of the two disconnected replicas (with the subset $C=\varnothing$) and $Z(\lambda=1)$ is the replicated partition function where the entanglement region $A$ is fully connected in imaginary time direction (with $C=A$). 
Now, we would like to obtain $\frac{Z(\lambda_{k+1})}{Z(\lambda_{k})}$ in PQMC according to Eq.~\eqref{eq:eqM18}, or more explicitly, 
\begin{equation}\label{eq:zoz}
\frac{Z(\lambda_{k+1})}{Z(\lambda_{k})} = \sum_{s_1,s_2,C\in A} \mathbf{P}_{s_1,s_2,C}(\lambda_k) O_C(\lambda_k),
\end{equation}
where 
\begin{equation}
\mathbf{P}_{s_1,s_2,C}(\lambda_k) = \frac{\lambda_k^{N_C}\left(1-\lambda_k\right)^{N_A-N_C} W_{s_1} W_{s_1} \operatorname{det} g_C^{s_1, s_2} }{ \sum_{s_1,s_2,C\in A} \lambda_k^{N_C}\left(1-\lambda_k\right)^{N_A-N_C} W_{s_1} W_{s_1} \operatorname{det} g_C^{s_1, s_2} }  
\end{equation}
and $O_C(\lambda_k) = \left(\frac{\lambda_{k-1}}{\lambda_{k}}\right)^{N_C}\left(\frac{1-\lambda_{k-1}}{1-\lambda_{k}}\right)^{N_A-N_C}$.
Then we need to update configurations $\{s_1\}$, $\{s_2\}$ and $\{C\}$.

The rest of this supplementary material is organized as follows: in Sec.~\ref{sec:workflow}, we present the workflow of the incremental algorithm. Then  we introduce how to update $\{s_1\}$ in Sec.~\ref{sec:s1}, $\{s_2\}$ in Sec.~\ref{sec:s2} and $\{C\}$ in Sec.~\ref{sec:c}.

\section{Workflow of the incremental algorithm}\label{sec:workflow}
The flow diagram of our incremental algorithm is shown in Fig.~\ref{fig:figS1}, the overbar in Fig.~\ref{fig:figS1} stands for the parallel pieces of $\lambda\in[0,1]$ and for one interval $\left[\lambda_k,\lambda_{k+1}\right]$. Below we discuss the steps in the order as in the figure.

\begin{enumerate}
\item Preparing the initial random configurations \{$s_1,s_2,C$\}, then calculating the equal-time Green's function $G^{s_1}(\Theta)$, $G^{s_2}(\Theta)$, the time displaced Green's function $G^{s_1}(\tau,\Theta)$, $G^{s_1}(\Theta,\tau)$,  $G^{s_2}(\tau,\Theta)$, $G^{s_2}(\Theta,\tau)$ and the inverse of Grover matrix $\left[g_C^{s_1, s_2}\right]^{}$ in terms of initial random configurations.

\item Sweeping $\tau$ from $\Theta$ to $2\Theta$.

2a. Trying to update $s_1$ according to Sec.~\ref{sec:s1}.

2b. Trying to update $s_2$ according to Sec.~\ref{sec:s2}.

2c. Trying to update $C$ according to Sec.~\ref{sec:c}, then measuring $O_C(\lambda_k) $. 

\item Sweeping $\tau$ from $2\Theta$ to $\Theta$.

\item Sweeping $\tau$ from $\Theta$ to $0$.

\item Sweeping $\tau$ from $0$ to $\Theta$. 

\item Repeating 2, 3, 4 and 5.

\end{enumerate}

We note that, to avoid brute-forcely invert a matrix,  the  Sherman-Morison formula
\begin{equation}
(\mathbf{A}_{n\times n}+\mathbf{U}_{n\times m} \mathbf{V}_{m\times n})^{-1}=\mathbf{A}^{-1}_{n\times n}-\mathbf{A}^{-1}_{n\times n}\mathbf{U}_{n\times m}\left(\mathbb{I}_m+\mathbf{V}_{m\times n} \mathbf{A}^{-1}_{n\times n}\mathbf{U}_{n\times m}\right)^{-1} \mathbf{V}_{m\times n} \mathbf{A}^{-1}_{n\times n},
\end{equation}
where $\mathbb{A}$ is $n\times n$ matrix, $\mathbf{U}$ is $n\times m$ matrix and $\mathbf{V}$ is $m\times n$ matrix, has been used in many places during in the workflow. For example, in simulation of Hubbard model, we have $n=N, m=1$. The formula turns the inversion of a rank-$n$ correction of a matrix (for example, the update of the Green's function when one auxiliary field at site $i$ and time $\tau$ is accepted) into the rank-$n$ correction to the inversion of the original matrix.
\begin{figure}[htp!]
	\centering
	\includegraphics[width=0.8\columnwidth]{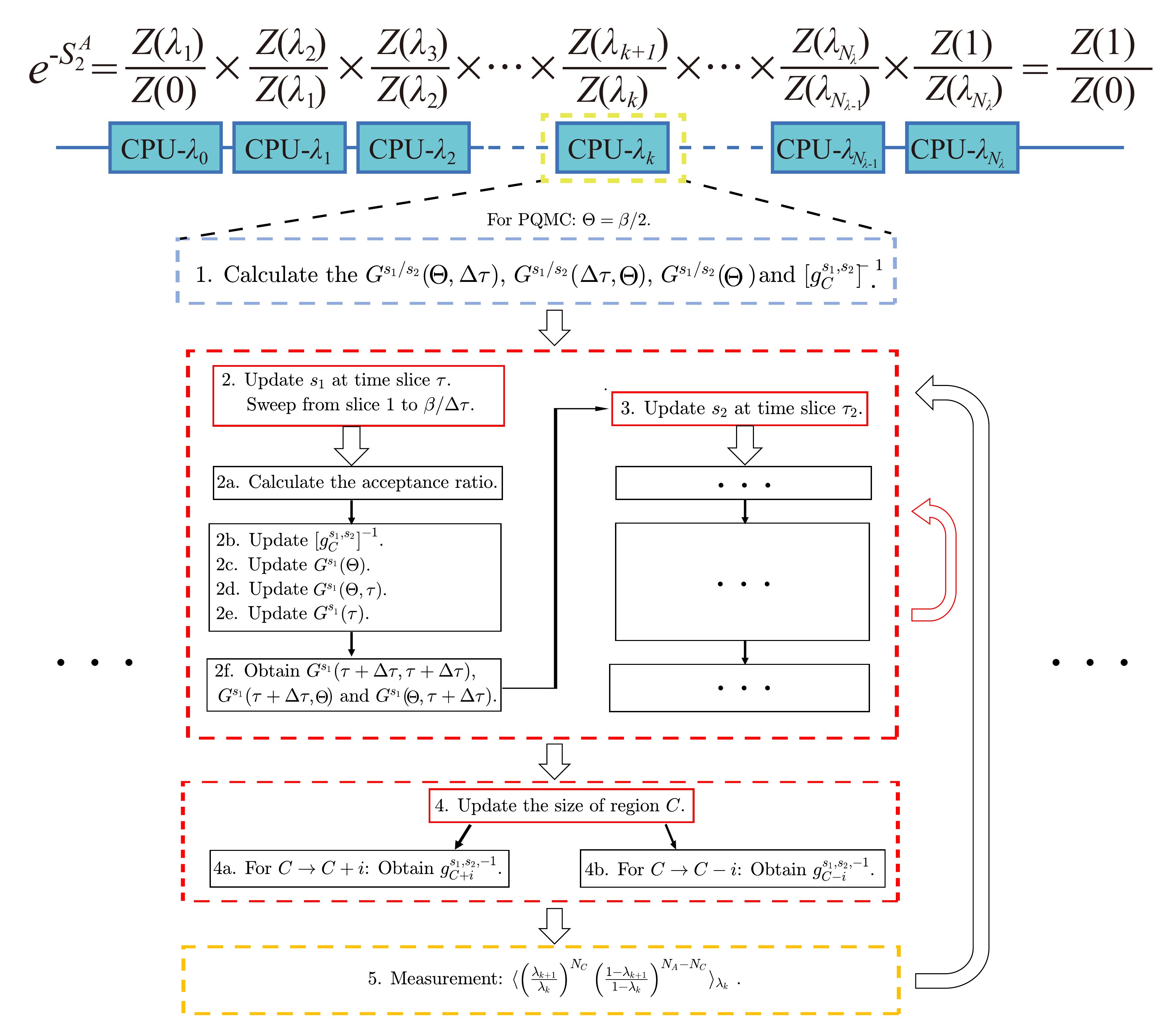}
	\caption{{The flow diagram of the incremental algorithm.} The algorithm splits a consecutive $\lambda-$parameterized equilibrium process into $N_{\lambda}$ independent smaller pieces. For each piece we start from a thermalized state of the partition function $Z(\lambda=\lambda_k)$ and carry out the equilibrium measurement which is determined by $\lambda_k$ and $\lambda_{k+1}$. For the $\lambda=0$ piece (the left column), the starting configuration is two independent replicas with ordinary periodical conditions and as the system evolves to $\lambda=\lambda_k$ the configuration becomes two modified replicas with some sites in region $A$ are glued together. The $\frac{Z(\lambda_{k-1})}{Z(\lambda_{k})}$  and $\frac{Z(\lambda_{k})}{Z(\lambda_{k+1})}$ pieces are carried in parallel. The final entanglement entropy is obtained from the summation of these independent pieces, as shown in Eq. (18) in the Method section. }
	\label{fig:figS1}
\end{figure}

\section{Updating \{$s_1$\} } \label{sec:s1}
Following the standard PQMC technicals introduced in Ref.~\cite{assaadWorld-line2008}, we define $\Delta=e^{V(s^\prime_1)} e^{-V (s_1)}-\mathbb{I}$, where $V(s_{1}) = \alpha \eta(s_1) n_\square$ for plaquette Hubbard model, thus $\Delta$ is a $4 \times 4$ matrix. While $\Delta$ is a number for on-site Hubbard model.
According to Eq.~\eqref{eq:zoz}, acceptance or rejection of this update of $s_1$ is up to the accept ratio,
\begin{equation}
R=\frac{W_{s_1^{\prime}}}{W_{s_1}} \frac{\operatorname{det} g_C^{s_1^{\prime}, s_2} }{\operatorname{det} g_C^{s_1, s_2} }.
\end{equation}
In PQMC, 
\begin{equation}
\begin{aligned}
\frac{W_{s_1^{\prime}}}{W_{s_1}}=& \frac{\operatorname{det}[B^{\langle}_\tau \left(\mathbb{I}+ {\Delta}\right) B^{\rangle}_\tau]}{\operatorname{det}[B^{\langle}_\tau B^{\rangle}_\tau]} \\
%=& \operatorname{det}[\mathbf{B}^{<}(\tau)\left(\mathbb{I}+\boldsymbol{\Delta}\right) \mathbf{B}^{>}(\tau)\left(\mathbf{B}^{<}(\tau)\mathbf{B}^{>}(\tau)\right)^{-1}] \\
=& \operatorname{det}[\mathbb{I}+{\Delta}(\mathbb{I}-{G}^{s_1}(\tau))].
\end{aligned}\label{eq:r0}
\end{equation}
We further define $\mathbf{r} = \mathbb{I}+{\Delta}(\mathbb{I}-{G}^{s_1}(\tau))$, and we could write down some useful equations
\begin{gather}
G^{s_1^\prime}(\tau)=G^{s_1}(\tau) - G^{s_1}(\tau) \mathbf{r}^{-1} {\Delta} [\mathbb{I}-G^{s_1}(\tau)] ,\\
G^{s_1^\prime}(\Theta)=G^{s_1}(\Theta)+G^{s_1}(\Theta,\tau) \mathbf{r}^{-1} {\Delta} G^{s_1}(\tau,\Theta) ,\\
G^{s_1^\prime}(\tau, \Theta)=G^{s_1}(\tau, \Theta)+G^{s_1}(\tau) \mathbf{r}^{-1} \Delta G^{s_1}(\tau, \Theta) ,\\
G^{s_1^\prime}(\Theta, \tau)=G^{s_1}(\Theta, \tau)-G^{s_1}(\Theta, \tau) \mathbf{r}^{-1} \Delta[\mathbb{I}-G^{s_1}(\tau)].
\end{gather} 
As for the part $\frac{\operatorname{det} g_C^{s_1^{\prime}, s_2}}{\operatorname{det} g_C^{s_1, s_2} }$, first we rewrite the Grover matrix as
\begin{equation}
g_C^{s_1, s_2}=G_C^{s_1}(\Theta) G_C^{s_2}(\Theta)+\left[\mathbb{I}-G_C^{s_1}(\Theta)\right]\left[\mathbb{I}-G_C^{s_2}(\Theta)\right]=\left[\mathbb{I}-G_C^{s_2}(\Theta)\right]+ G_C^{s_1}(\Theta) \left[2 G_C^{s_2}(\Theta)-\mathbb{I}\right].
\end{equation}
Here we strictly define the Grover matrix at time slice $\tau=\Theta$.
Then
\begin{equation}
\begin{aligned}
\frac{g_C^{s_1^{\prime}, s_2}}{g_C^{s_1, s_2}} &=\frac{\left[\mathbb{I}-G_C^{s_2}(\Theta)\right]+G_C^{s'_1}(\Theta)\left[2 G_C^{s_2}(\Theta)-\mathbb{I}\right]}{\left[\mathbb{I}-G_C^{s_2}(\Theta)\right]+G_C^{s_1}(\Theta)\left[2 G_C^{s_2}(\Theta)-\mathbb{I}\right]}\\
&=\frac{\left[\mathbb{I}-G_C^{s_2}(\Theta)\right]+G_C^{s_1}(\Theta)\left[2 G_C^{s_2}(\Theta)-\mathbb{I}\right]+\left[G^{s_1}(\Theta,\tau) \mathbf{r}^{-1} \Delta G^{s_1}(\tau,\Theta)\right]_{C} \left[2 G_C^{s_2}(\Theta)-\mathbb{I}\right] }{\left[\mathbb{I}-G_C^{s_2}(\Theta)\right]+G_C^{s_1}(\Theta)\left[2 G_C^{s_2}(\Theta)-\mathbb{I}\right]}\\
&=\mathbb{I}+\left[G^{s_1}(\Theta,\tau) \mathbf{r}^{-1} \Delta G^{s_1}(\tau,\Theta)\right]_C \left[2 G_C^{s_2}(\Theta)-\mathbb{I}\right][g_C^{s_1, s_2}]^{-1},
\end{aligned}
\end{equation}
where subscript $C$ means the corresponding matrix is $N_C\times N_C$.
Thus 
\begin{equation}
\begin{aligned}
\frac{\operatorname{det} g_C^{s_1^{\prime}, s_2} }{\operatorname{det} g_C^{s_1, s_2}} &=\det [\mathbb{I}+\left[G^{s_1}(\Theta,\tau) \mathbf{r}^{-1} \Delta G^{s_1}(\tau,\Theta)\right]_C \left[2 G_C^{s_2}(\Theta)-\mathbb{I}\right][g_C^{s_1, s_2}]^{-1}]\\
&=\det[\mathbb{I}+\mathbf{r}^{-1} \Delta G^{s_1}(\tau,\Theta) \left[2 G_C^{s_2}(\Theta)-\mathbb{I}\right][g_C^{s_1, s_2}]^{-1} G^{s_1}(\Theta,\tau)  ],
\end{aligned}\label{eq:r1}
\end{equation}
where the Sylvester's determinant theorem is used. And we could notice that we need to store $[g_C^{s_1, s_2}]^{-1}$ in code instead of $g_C^{s_1, s_2}$.

We employ the Metropolis algorithm $r > \min\{1,R\}$, here $r\in (0,1]$ is a random number, to update. 
If $s_1 \rightarrow s_1^\prime$ is accepted, we need to update $G^{s_1}(\tau)$, $G^{s_1}(\Theta)$, $G^{s_1}(\tau, \Theta)$, $G^{s_1}(\Theta,\tau)$ and $[g_C^{s_1, s_2}]^{-1}$.
$[g_C^{s_1, s_2}]^{-1}$ should be updated by
\begin{equation}
\begin{aligned}
&[\frac{g_C^{s_1^{\prime}, s_2}}{g_C^{s_1, s_2}}]^{-1} = g_C^{s_1, s_2} [g_C^{s_1^{\prime}, s_2}]^{-1} =\left(\mathbb{I}+G^{s_1}(\Theta,\tau) \mathbf{r}^{-1} \Delta G^{s_1}(\tau,\Theta) \left[2 G_C^{s_2}(\Theta)-\mathbb{I}\right][g_C^{s_1, s_2}]^{-1}\right)^{-1}\\
&=\mathbb{I}-G^{s_1}(\Theta,\tau) \left(\mathbb{I}+\mathbf{r}^{-1} \Delta G^{s_1}(\tau,\Theta) \left[2 G_C^{s_2}(\Theta)-\mathbb{I}\right][g_C^{s_1, s_2}]^{-1} G^{s_1}(\Theta,\tau)  \right)^{-1} \mathbf{r}^{-1} \Delta G^{s_1}(\tau,\Theta) \left[2 G_C^{s_2}(\Theta)-\mathbb{I}\right][g_C^{s_1, s_2}]^{-1}
\end{aligned}\label{eq:g1}
\end{equation}
where the Sherman-Morison formula is used.

\section{Updating $\{s_2\}$} \label{sec:s2}
In this case, the accept ratio 
\begin{equation}
R=\frac{W_{s_2^{\prime}}}{W_{s_2}} \frac{\operatorname{det} g_C^{s_1, s_2^{\prime}} }{\operatorname{det} g_C^{s_1, s_2} },
\end{equation}
where $W_{s_2^{\prime}}/W_{s_2} = \operatorname{det}[\mathbb{I}+{\Delta}(\mathbb{I}-{G}^{s_2}(\tau))]$.
We firstly rewrite the Grover matrix
\begin{equation}
g_C^{s_1, s_2}=G_C^{s_1}(\Theta) G_C^{s_2}(\Theta)+\left[\mathbb{I}-G_C^{s_1}(\Theta)\right]\left[\mathbb{I}-G_C^{s_2}(\Theta)\right]=\left[\mathbb{I}-G_C^{s_1}(\Theta)\right]+\left[2 G_C^{s_1}(\Theta)-\mathbb{I}\right]G_C^{s_2}(\Theta),
\end{equation}
then we could easily calculate $\frac{\operatorname{det} g_C^{s_1, s_2^{\prime}}}{\operatorname{det} g_C^{s_1, s_2} }$, and update Green's functions and the inverse of Grover matrix according to similar procedure as Sec.~\ref{sec:s1}.

\section{Updating $\{C\}$}\label{sec:c}
For region $C \rightarrow C + i$ , the accept ratio is given as
\begin{equation}
R_{\text {plus }}=\frac{\lambda_k}{1-\lambda_k} \frac{\operatorname{det} g_{C+i}^{s_1, s_2} }{\operatorname{det} g_C^{s_1, s_2} }.
\end{equation}
We explicitly write $g_C^{s_1, s_2}=\mathbb{I}+2 G_C^{s_1}(\Theta) G_C^{s_2}(\Theta)-G_C^{s_1}(\Theta)-G_C^{s_2}(\Theta)$, then omitting $\Theta$ for conciseness,
\begin{equation}
\begin{aligned}
& g_{C+i}^{s_1, s_2} =\left(\begin{array}{cc}
I_{C C} & 0 \\
0 & I_{i i}
\end{array}\right)+2\left(\begin{array}{cc}
G_{C C}^{s_1} & G_{C i}^{s_1} \\
G_{i C}^{s_1} & G_{i i}^{s_1}
\end{array}\right)\left(\begin{array}{cc}
G_{C C}^{s_2} & G_{C i}^{s_2} \\
G_{i C}^{s_2} & G_{i i}^{s_2}
\end{array}\right)-\left(\begin{array}{cc}
G_{C C}^{s_1} & G_{C i}^{s_1} \\
G_{i C}^{s_1} & G_{i i}^{s_1}
\end{array}\right)-\left(\begin{array}{cc}
G_{C C}^{s_2} & G_{C i}^{s_2} \\
G_{i C}^{s_2} & G_{i i}^{s_2}
\end{array}\right) \\
%& =\left(\begin{array}{cc}
%I_{C C}+2 G_{C C}^{s_1} G_{C C}^{s_2}-G_{C C}^{s_1}-G_{C C}^{s_2}+2 G_{C i}^{s_1} G_{i C}^{s_2} & 2 G_{C C}^{s_1} G_{C i}^{s_2}-G_{C i}^{s_1}-G_{C i}^{s_2}+2 G_{C i}^{s_1} G_{i i}^{s_2} \\
%2 G_{i C}^{s_1} G_{C C}^{s_2}-G_{i C}^{s_1}-G_{i C}^{s_2}+2 G_{i i}^{s_1} G_{i C}^{s_2} & I_{i i}+2 G_{i C}^{s_1} G_{C i}^{s_2}-G_{i i}^{s_1}-G_{i i}^{s_2}+2 G_{i i}^{s_1} G_{i i}^{s_2}
%\end{array}\right) \\
& =\left(\begin{array}{cc}
g_C^{s_1, s_2}+2 G_{C i}^{s_1} G_{i C}^{s_2} & 2 G_{C C}^{s_1} G_{C i}^{s_2}-G_{C i}^{s_1}-G_{C i}^{s_2}+2 G_{C i}^{s_1} G_{i i}^{s_2} \\
2 G_{i C}^{s_1} G_{C C}^{s_2}-G_{i C}^{s_1}-G_{i C}^{s_2}+2 G_{i i}^{s_1} G_{i C}^{s_2} & I_{i i}+2 G_{i C}^{s_1} G_{C i}^{s_2}-G_{i i}^{s_1}-G_{i i}^{s_2}+2 G_{i i}^{s_1} G_{i i}^{s_2}
\end{array}\right) \\
& \equiv\left(\begin{array}{cc}
A & B \\
C & D
\end{array}\right).
\end{aligned}
\end{equation}
We notice that $D$ is a number or $1\times1$-matrix, then we have
\begin{equation}
\det g_{C+i}^{s_1, s_2}  = \det\left(\begin{array}{ll}
A & B \\
C & D
\end{array}\right)=\det\left(A\right) \left(D-C A^{-1} B\right).
\end{equation}
We could obtain $A^{-1}$ by
\begin{equation}
\begin{aligned}
A^{-1}&=\left[g_C^{s_1, s_2}+2 G_{C i}^{s_1} G_{i C}^{s_2}\right]^{-1}\\
&=g_C^{-1,s_1, s_2}-2 g_C^{-1,s_1, s_2} G_{C i}^{s_1}\left(\mathbb{I}+2G_{i C}^{s_2}g_C^{-1,s_1, s_2}G_{C i}^{s_1}\right)^{-1} G_{i C}^{s_2}g_C^{-1,s_1, s_2}
\end{aligned}
\end{equation}
and components B, C and D could be obtained directly by $G^{s_1}(\Theta)$ and $G^{s_2}(\Theta)$.
Then
\begin{equation}
\begin{aligned}
\frac{\operatorname{det} g_{C+i}^{s_1, s_2} }{\operatorname{det} g_C^{s_1, s_2} } &=\det\frac{A}{g_C^{s_1, s_2}} \left(D-C A^{-1} B\right) \\
&=\det\left(\mathbb{I}+2 G_{C i}^{s_1} G_{i C}^{s_2}g_C^{-1,s_1, s_2}\right) \left(D-C A^{-1} B\right)  \\
&=\left(1+2  G_{i C}^{s_2}g_C^{-1,s_1, s_2}G_{C i}^{s_1}\right) \left(D-C A^{-1} B\right)
\end{aligned}\label{eq:r3}
\end{equation} 
If $P_{\text{plus}} = \min\left\{1,R_{\text{plus}} \right\} $ is accepted, we could further introduce
\begin{equation}
[g_{C+i}^{s_1, s_2}]^{-1} = \left[\begin{array}{ll}
A & B \\
C & D
\end{array}\right]^{-1}=\left[\begin{array}{cc}
A^{-1}+A^{-1} B\left(D-C A^{-1} B\right)^{-1} C A^{-1} & -A^{-1} B\left(D-C A^{-1} B\right)^{-1} \\
-\left(D-C A^{-1} B\right)^{-1} C A^{-1} & \left(D-C A^{-1} B\right)^{-1}
\end{array}\right],
\end{equation}

While for region $C \rightarrow C - i$ ,  we note $\tilde{C} = C - i$, then 
\begin{equation}
R_{\text {minus }}=\frac{1-\lambda_k}{\lambda_k} \frac{\operatorname{det}\left(g_{\tilde{C}}^{s_1, s_2}\right)}{\operatorname{det}\left(g_C^{s_1, s_2}\right)}
\end{equation}
and explicitly
\begin{equation}
\begin{aligned}
& g_C^{s_1, s_2} =\left(\begin{array}{cc}
I_{\tilde{C} \tilde{C}} & 0 \\
0 & I_{i i}
\end{array}\right)+2\left(\begin{array}{cc}
G_{\tilde{C} \tilde{C}}^{s_1} & G_{\tilde{C} i}^{s_1} \\
G_{i \tilde{C}}^{s_1} & G_{i i}^{s_1}
\end{array}\right)\left(\begin{array}{cc}
G_{\tilde{C} \tilde{C}}^{s_2} & G_{\tilde{C} i}^{s_2} \\
G_{i \tilde{C}}^{s_2} & G_{i i}^{s_2}
\end{array}\right)-\left(\begin{array}{cc}
G_{\tilde{C}}^{s_1} \tilde{C} & G_{\tilde{C} i}^{s_1} \\
G_{i \tilde{C}}^{s_1} & G_{i i}^{s_1}
\end{array}\right)-\left(\begin{array}{cc}
G_{\tilde{C}}^{s_2} \tilde{C} & G_{\tilde{C} i}^{s_2} \\
G_{i \tilde{C}}^{s_2} & G_{i i}^{s_2}
\end{array}\right) \\
%& =\left(\begin{array}{cc}
%I_{\tilde{C} \tilde{C}}+2 G_{\tilde{C} \tilde{C}}^{s_1} G_{\tilde{C} \tilde{C}}^{s_2}-G_{\tilde{C} \tilde{C}}^{s_1}-G_{\tilde{C} \tilde{C}}^{s_2}+2 G_{\tilde{C} i}^{s_1} G_{i \tilde{C}}^{s_2} & 2 G_{\tilde{C} \tilde{C}}^{s_1} G_{\tilde{C} i}^{s_2}-G_{\tilde{C} i}^{s_1}-G_{\tilde{C} i}^{s_2}+2 G_{\tilde{C} i}^{s_{\tilde{C}}} G_{i i}^{s_2} \\
%2 G_{i \tilde{C}}^{s_1} G_{\tilde{C} \tilde{C}}^{s_2}-G_{i \tilde{C}}^{s_1}-G_{i \tilde{C}}^{s_2}+2 G_{i i}^{s_1} G_{i \tilde{C}}^{s_2} & I_{i i}+2 G_{i \tilde{C}}^{s_1} G_{\tilde{C} i}^{s_2}-G_{i i}^{s_1}-G_{i i}^{s_2}+2 G_{i i}^{s_1} G_{i i}^{s_2}
%\end{array}\right) \\
& =\left(\begin{array}{cc}
g_{\tilde{C}}^{s_1, s_2}+2 G_{\tilde{C} i}^{s_1} G_{i \tilde{C}}^{s_2} & 2 G_{\tilde{C} \tilde{C}}^{s_1} G_{\tilde{C} i}^{s_2}-G_{\tilde{C} i}^{s_1}-G_{\tilde{C} i}^{s_2}+2 G_{\tilde{C} i}^{s_{\tilde{C}}} G_{i i}^{s_2} \\
2 G_{i \tilde{C}}^{s_1} G_{\tilde{C} \tilde{C}}^{s_2}-G_{i \tilde{C}}^{s_1}-G_{i \tilde{C}}^{s_2}+2 G_{i i}^{s_1} G_{i \tilde{C}}^{s_2} & I_{i i}+2 G_{i \tilde{C}}^{s_1} G_{\tilde{C} i}^{s_2}-G_{i i}^{s_1}-G_{i i}^{s_2}+2 G_{i i}^{s_1} G_{i i}^{s_2}
\end{array}\right) \\
& \equiv\left(\begin{array}{ll}
\tilde{A} & \tilde{B} \\
\tilde{C} & \tilde{D}
\end{array}\right) \\
&
\end{aligned}
\end{equation}
and $\det g_C^{s_1, s_2} = \det \tilde{A} \left(\tilde{D}-\tilde{C} \tilde{A}^{-1} \tilde{B}\right)$.
Since
\begin{equation}
\begin{aligned}
& g_C^{-1, s_1, s_2}=\left(\begin{array}{cc}
\tilde{A} & \tilde{B} \\
\tilde{C} & \tilde{D}
\end{array}\right)^{-1} \\
& =\left(\begin{array}{cc}
\tilde{A}^{-1}+\tilde{A}^{-1} \tilde{B}\left(\tilde{D}-\tilde{C} \tilde{A}^{-1} \tilde{B}\right)^{-1} \tilde{C} \tilde{A}^{-1} & -\tilde{A}^{-1} \tilde{B}\left(\tilde{D}-\tilde{C} \tilde{A}^{-1} \tilde{B}\right)^{-1} \\
-\left(\tilde{D}-\tilde{C} \tilde{A}^{-1} \tilde{B}\right)^{-1} \tilde{C} \tilde{A}^{-1} & \left(\tilde{D}-\tilde{C} \tilde{A}^{-1} \tilde{B}\right)^{-1}
\end{array}\right)\\
&=\left(\begin{array}{ll}
g_{\tilde{C} \tilde{C}}^{-1, s_1, s_2} & g_{\tilde{C}, i}^{-1, s_1, s_2} \\
g_{i \tilde{C}}^{-1, s_1, s_2} & g_{i i}^{-1, s_1, s_2}
\end{array}\right)\\
&\equiv\left(\begin{array}{ll}
E & F \\
H & J
\end{array}\right) 
\end{aligned}
\end{equation}
Then
\begin{equation}
\begin{aligned}
\tilde{A}^{-1}&=E-\tilde{A}^{-1} \tilde{B}\left(\tilde{D}-\tilde{C} \tilde{A}^{-1} \tilde{B}\right)^{-1} \tilde{C} \tilde{A}^{-1}=E-F J^{-1}H
\end{aligned}
\end{equation}

Accept ratio is 
\begin{equation}
\begin{aligned}
\frac{\operatorname{det}  g_{\tilde{C}}^{s_1, s_2} }{\operatorname{det} g_C^{s_1, s_2}} &=\frac{\operatorname{det}\left(\tilde{A}- 2 G_{\tilde{C} i}^{s_1} G_{i \tilde{C}}^{s_2}\right)}{\det\left(\tilde{A}\right)} \left(\tilde{D}-\tilde{C} \tilde{A}^{-1} \tilde{B}\right)^{-1} \\
&=\det\left(\mathbb{I}-2 G_{\tilde{C} i}^{s_1} G_{i \tilde{C}}^{s_2}\tilde{A}^{-1}\right)  \left(g^{-1, s_1, s_2}\right)_{i i}  \\
&=\left(1-2 G_{i \tilde{C}}^{s_2}\tilde{A}^{-1}G_{\tilde{C} i}^{s_1}\right)\left(g^{-1, s_1, s_2}\right)_{i i} 
\end{aligned}\label{eq:r4}
\end{equation}

Then update inverse of Grover matrix:
\begin{equation}
\begin{aligned}
g_{\tilde{C}}^{-1,s_1,s_2}&=\left[\tilde{A}-2 G_{\tilde{C} i}^{s_1} G_{i \tilde{C}}^{s_2}\right]^{-1}\\
&=\tilde{A}^{-1}\left[\mathbb{I}-2 G_{\tilde{C} i}^{s_1} G_{i \tilde{C}}^{s_2}\tilde{A}^{-1}\right]^{-1}\\
&=\tilde{A}^{-1}\left[\mathbb{I}+2 G_{\tilde{C} i}^{s_1}\left(\mathbb{I}-2G_{i \tilde{C}}^{s_2}\tilde{A}^{-1}G_{\tilde{C} i}^{s_1}\right)^{-1} G_{i \tilde{C}}^{s_2}\tilde{A}^{-1}\right]\\
&=\tilde{A}^{-1}+2 \tilde{A}^{-1} G_{\tilde{C} i}^{s_1}\left(1-2G_{i \tilde{C}}^{s_2}\tilde{A}^{-1}G_{\tilde{C} i}^{s_1}\right)^{-1} G_{i \tilde{C}}^{s_2}\tilde{A}^{-1}
\end{aligned}\label{eq:g4}
\end{equation}

\end{document}